\documentclass[aps,prd,twocolumn,showpacs,preprintnumbers, nofootinbib]{revtex4}

\usepackage{natbib}
\usepackage{graphicx}
\usepackage[english]{babel}
\usepackage{color}
\usepackage{amsmath,amssymb,amsfonts,dsfont}
\usepackage{multirow}
\usepackage{sistyle}
\usepackage{ulem}

\graphicspath{{./figures/}}

\newcommand{\ds}{{\sf DarkSUSY}}

\newcommand{\be}{\begin{equation}}
\newcommand{\ee}{\end{equation}}
\newcommand{\bea}{\begin{eqnarray}}
\newcommand{\eea}{\end{eqnarray}}

\DeclareMathOperator{\dif}{d\!}
\DeclareMathOperator{\e}{\mathrm e}
\def\ket#1{|#1\rangle}
\def\bra#1{\langle #1|}
\def\bracket#1#2{\langle #1|#2\rangle}
\def\bvec#1{\textrm{\boldmath $#1 $}}

\begin{document}
\title{The 130\,GeV gamma-ray line and generic dark matter model building constraints from continuum gamma rays, radio and antiproton data}

\author{Masaki Asano}
\email{masaki.asano@desy.de}
\affiliation{{II.} Institute for Theoretical Physics, University of Hamburg, 
Luruper Chausse 149, DE-22761 Hamburg, Germany}

\author{Torsten~Bringmann}
\email{torsten.bringmann@desy.de}
\affiliation{{II.} Institute for Theoretical Physics, University of Hamburg, 
Luruper Chausse 149, DE-22761 Hamburg, Germany}

\author{G\"unter~Sigl}
\email{guenter.sigl@desy.de}
\affiliation{{II.} Institute for Theoretical Physics, University of Hamburg, 
Luruper Chausse 149, DE-22761 Hamburg, Germany}

\author{Martin Vollmann}
\email{martin.vollmann@desy.de}
\affiliation{{II.} Institute for Theoretical Physics, University of Hamburg, 
Luruper Chausse 149, DE-22761 Hamburg, Germany}

\date{16 May 2013}

\begin{abstract}
An analysis of the Fermi gamma ray space telescope data has recently
revealed a resolved gamma-ray feature close to the galactic center
which is consistent with monochromatic photons at an energy of about
130 GeV. If interpreted in terms of dark matter (DM) annihilating into
$\gamma\gamma$ ($\gamma Z$, $\gamma h$), this would correspond to a DM
particle mass of roughly 130\,GeV (145\,GeV, 155\,GeV). The rate for
these loop-suppressed processes, however, is  larger than typically
expected for thermally produced DM. Correspondingly, one would
generically expect even larger tree level production rates of standard
model fermions or gauge bosons. Here, we quantify this expectation in
a rather model-independent way by relating the tree level and loop
amplitudes with the help of the optical theorem. As an application, we
consider bounds from continuum gamma rays, radio and antiproton data
on the tree level amplitudes and translate them into constraints
on the loop amplitudes. We find that, independently of the DM
production mechanism, any DM model aiming at explaining the line
signal in terms of charged standard model particles running in the
loop is in rather strong tension with at least one of these
constraints, with the exception of loops dominated by top quarks. 
We stress that attempts to explain the 130\,GeV feature with internal bremsstrahlung 
do not suffer from such difficulties.

\end{abstract}


\maketitle


\begin{figure*}[t!]
 \begin{center}
        \includegraphics[width=0.65\textwidth]{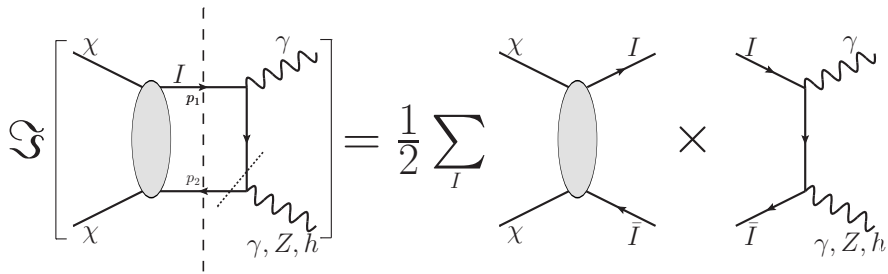} 
 \end{center}
\caption{\label{fig:OT} Graphical illustration of Eq.~(\ref{eq:unitarity3}), i.e.~the {\it optical theorem} applied to the case of DM annihilation into $\gamma X$, where $X=\gamma, Z, h$: all intermediate-state particles $I$ that can be put on-shell contribute to 
the imaginary part of the loop-amplitude. This allows to relate the latter to the tree-level annihilation rate of DM  into SM particles with $m_{\rm SM}\!<\!m_\chi$, where the corresponding $u$-channel diagram (as well as the 4-point interaction for $W^+W^-\rightarrow\gamma X$) is not shown explicitly in the figure. For $\gamma Z$ and $\gamma h$ final states dominated by fermion loops, another possible way of putting the intermediate particles on-shell -- compared to what is shown on the right hand side -- results from 'cutting' the loop diagram along the dotted rather than  the dashed lines. }
\end{figure*}

\section{Introduction}

A very promising way to search for dark matter (DM) particles is to
look for the pronounced spectral signatures in cosmic gamma rays which
are expected from DM annihilation or decay
\cite{Bringmann:2011ye}. A (quasi-) monochromatic gamma-ray line, in
particular, results if DM annihilates or decays into final states
including a photon (such as $\gamma\gamma$, $\gamma Z$ or $\gamma h$)
\cite{Bergstrom:1997fj} or, in particular for Majorana DM
particles, due to internal bremsstrahlung (IB) photons for $f\bar
f\gamma$ final states \cite{Bringmann:2007nk}. Such a feature is
usually considered a {\it smoking gun} signature for particle DM as
there are no known astrophysical processes that  produce
mono-energetic gamma rays at high energies.

The recent discovery \cite{Bringmann:2012vr,Weniger:2012tx} of a tentative line-like spectral feature in the data of the Fermi Gamma-Ray Space Telescope \cite{Atwood:2009ez} (independently confirmed in Ref.~\cite{Tempel:2012ey}) has therefore prompted considerable interest in the field, in particular after a  sophisticated spatial template analysis by Su \& Finkbeiner has confirmed its existence with an impressive {\it global} significance of slightly more than $5\,\sigma$ \cite{Su:2012ft}. 
Instrumental systematics seem unlikely as a cause for this gamma-ray line at an energy of around 130\,GeV \cite{Finkbeiner:2012ez} and no compelling alternative explanation to DM annihilation 
has been brought forward so far (but see Ref.~\cite{Aharonian:2012cs}). 
However,  there also exist a few indications which eventually may, if confirmed, disfavor the DM interpretation, so more data is certainly needed to settle this issue. For more details about the current status of the signal and possible implications for DM models, including an extensive list of references, we refer the reader to a dedicated review \cite{line_review}. 

One possible worry is that the required annihilation rate is
considerably larger than typically expected for thermally produced DM \cite{Lee:1977ua,Griest:1990kh},
which means that one may have to abandon this theoretically very
appealing mechanism for DM production in the early universe. Even in that case, however, a large annihilation rate into $\gamma\gamma$, $\gamma Z$ or $\gamma h$ final states usually implies a much larger annihilation rate of DM particles into pairs of standard model (SM) fermions or weak gauge bosons -- simply because the former processes can only happen at the 1-loop level, while the latter happen at tree level. It has been noted that these annihilation products could therefore rather easily violate existing bounds on continuum gamma rays \cite{Buckley:2012ws,Buchmuller:2012rc,Cohen:2012me,Cholis:2012fb} or cosmic-ray antiprotons \cite{Buchmuller:2012rc} in typical scenarios where DM consists of weakly interacting massive particles (WIMPs). In fact, it has already been claimed \cite{Cohen:2012me} that the leading DM candidate, the lightest neutralino in supersymmetric theories \cite{Jungman:1995df}, is ruled out as an explanation of the observed line signal (see, however, Refs.~\cite{line_review,Shakya:2012fj,Lee:2012bq,Kang:2012bq,Das:2012ys,Li:2012tra,SchmidtHoberg:2012ip}).

Here, we revisit these arguments in an as systematic and
model-independent way as possible. 
As sketched in Fig.~\ref{fig:OT}, we
use the optical theorem
to relate the expected annihilation rates at tree-level to the
imaginary part of the loop-level amplitude that is supposed to
account for the line signal  (extending, and in a certain sense
turning around, the recent analysis by Abazajian {\it et
  al}.~\cite{Abazajian:2011tk}). We then confront these tree-level
annihilation rates, on a channel-by-channel basis, to constraints
arising from the non-observation of excesses in continuum gamma
rays from the galactic center and dwarf galaxies, as well as
antiproton and radio data.

Our main result is that the expected large tree-level annihilation 
rates are indeed  both a generic and serious challenge to 
{\it any} DM model trying to explain the observed 130 GeV feature 
by a loop that involves SM particles. Given that  DM  
usually couples directly to SM particles in commonly adopted WIMP 
frameworks like supersymmetry, this has severe implications on 
concrete model building attempts.
We classify the (few) possible loopholes to the above statement and 
stress that IB (which upcoming experiments could distinguish 
sufficiently well from monochromatic photons \cite{Bergstrom:2012vd}) may 
well be a more likely explanation of the signal in terms of DM annihilation. 
Alternatively, our results seem to imply that the dominant coupling of 
DM to charged states must be exclusively to new particles which are 
heavier than the DM particles themselves -- a possibility which may not be
very common in typical WIMP scenarios (aiming simultaneously at addressing 
the fine-tuning issues of the SM) but which has seen a lot of interest
in response to the observed photon excess at 130 GeV 
(see, e.g., Refs.~\cite{Buckley:2012ws,Dudas:2012pb,Cline:2012nw,Choi:2012ap,Wang:2012ts,Chalons:2012xf}).

This article is organized as follows. We start in Section \ref{sec:OT} 
by a short review about the optical theorem and how one can use
the strength of the potentially observed line signal at 130\,GeV to  derive a lower 
bound on the expected DM annihilation rate at tree level. Constraints 
on tree-level annihilation rates from both continuum gamma-ray, antiproton 
and radio observations are then discussed in Section \ref{sec:constraints}. We 
combine these results in Section  \ref{sec:model} to constrain the 
maximally allowed fractional contribution of the imaginary part of the loop-amplitude to the full 
annihilation cross section into $\gamma X$ (with $X=\gamma, Z, h$) and discuss implications for 
DM model building. In Section \ref{sec:conc}, we present our conclusions and an outlook.
In the Appendix, we collect some technical details about our calculations
and the application of the optical theorem to the particular processes we are interested in here.
Throughout the paper we use natural units, $c_0=\hbar=1$, unless indicated explicitly.

\section{From loop to tree-level processes}
\label{sec:OT}

Our main goal is to establish a relation between the 1-loop level
cross sections $\sigma_{\gamma X}$ for dark matter annihilation into
$\gamma X$ final states and the
tree level cross sections $\sigma_I$ to pairs of Standard Model or other
particles which form intermediate states denoted by $I$ for the production of this final state. To this end, we express the unitarity, $S^\dagger
S=\mathds{1}$, of the $S$-matrix $S=\mathds{1}+iT$ in terms of the
transition matrix $T$,
\begin{equation}\label{eq:unitarity}
  -i(T-T^\dagger)=T^\dagger T\,.
\end{equation}
Sandwiching this equation between an initial and a final state
$|i\rangle$ and $|f\rangle$, respectively, yields
\begin{equation}\label{eq:unitarity2}
  \Im\left[\left\langle f|T|i\right\rangle\right]=\frac{1}{2}\sum_X
  \left\langle f|T^\dagger|X\right\rangle \left\langle X|T|i\right\rangle\,,
\end{equation}
where the sum runs over all on-shell intermediate states $|X\rangle$ whose total four
momenta equal the initial state four-momentum $p$,
in our case consisting of two dark matter particles $\chi$,
and whose total angular momentum equals the one of the initial and
final state $(J,M)$. Note that this assumes $\langle f|T^\dagger|i\rangle=\langle f|T|i\rangle^*$ -- as is, e.g., satisfied in the angular momentum basis if the theory is invariant under time-reversal
    \cite{Abazajian:2011tk}, which 
is the situation we are interested in here (the same would be true when using a helicity basis).
Obviously, $|X\rangle$ must also share all other conserved quantum numbers 
with $|i\rangle$ and $|f\rangle$, such as in particular parity.
We now assume that this sum only contains states
consisting of two free particles of type $I$. Exploiting energy
momentum conservation in the center of mass frame all relevant two
particle states can be
uniquely characterized by the total spin $S$ of the two particles, its
projection $m_S$ on some given coordinate axis, and the unit vector
$\hat{\bf p}_1$ of one of the particles. Thus for the angular part of
the wave functions for such states we can write
\begin{widetext}
\begin{equation}\label{eq:X_momentum}
  |X\rangle=|\hat{\bf p}_1,S,m_S\rangle_I=\sum_{L,m_L}{}_I\langle
  L,m_L,S,m_S|\hat{\bf p}_1,S,m_S\rangle_I|L,m_L,S,m_S\rangle_I=
  \sum_{L,m_L}Y_{L,m_L}(\hat{\bf p}_1)|L,m_L,S,m_S\rangle_I\,,
\end{equation}
\end{widetext}
where we have expanded this state in terms of the eigenstate
$|L,m_L,S,m_S\rangle_I$ of orbital angular momentum and spin and we have
used that the expansion coefficient is simply
the spherical harmonic function,
$_I\langle L,m_L,S,m_S|\hat{\bf p}_1,S,m_S\rangle_I=Y_{L,m_L}(\hat{\bf
  p}_1)$. Applying $\langle f|T^\dagger|$ from the left to
Eq.~(\ref{eq:X_momentum}) and inverting to $\langle
f|T^\dagger|L,m_L,S,m_S\rangle_I$ by using the orthonormality $\int
d^2\hat{\bf p}_1Y^*_{L_1,m_1}(\hat{\bf p}_1)Y_{L_2,m_2}(\hat{\bf
  p}_1)=\delta_{L_1,L_2}\delta_{m_1,m_2}$ gives
\begin{widetext}
\begin{eqnarray}\label{eq:X_momentum2}
  \langle f|T^\dagger|L,m_L,S,m_S\rangle_I&=&\int d^2\hat{\bf p}_1 Y_{L,m_L}(\hat{\bf
  p}_1)\langle f|T^\dagger|\hat{\bf p}_1,S,m_S\rangle_I\equiv\langle
  J,M|L,m_L,S,m_S\rangle_I{\cal M}^*_{f\to(I,L,S)}(s)\nonumber\\
  \langle f|T^\dagger|\hat{\bf p}_1,S,m_S\rangle_I&=&\sum_{L,m_L}
  Y^*_{L,m_L}(\hat{\bf p}_1)\langle J,M|L,m_L,S,m_S\rangle_I{\cal M}^*_{f\to(I,L,S)}(s)\,.
\end{eqnarray}
\end{widetext}
Here we have used the fact that if the transition matrix is a scalar
the amplitude $\langle f|T^\dagger|L,m_L,S,m_S\rangle_I$ has to be
proportional to the Clebsch-Gordan coefficient $\langle
J,M|L,m_L,S,m_S\rangle$ and we denoted by ${\cal
  M}_{i\to(I,L,S)}(s)$ the remaining factor which is
independent of the projections $m_S$ and $m_L$ and can only depend on
the total angular momenta $L$ and $S$ and on the Lorentz invariant
$s=p^2\equiv(p_1+p_2)^2$. Eq.~(\ref{eq:X_momentum2}) provides the relation between the
amplitude $\langle f|T^\dagger|\hat{\bf p}_1,S,m_S\rangle_I$ that is
calculated directly from the Feynman rules and the amplitudes ${\cal
  M}_{i\to(I,L,S)}(s)$ that are more convenient for the following
analysis. For Eq.~(\ref{eq:unitarity2}) we can now write
\begin{widetext}
\begin{equation}\label{eq:unitarity3}
  \Im\left[{\cal M}_{i\to f}(s)\right]=\frac{1}{2}\int\frac{1}{2E_1}\frac{d^3{\bf
      p_1}}{(2\pi)^3}\int\frac{1}{2E_2}\frac{d^3{\bf
      p_2}}{(2\pi)^3}\sum_I\langle f|T^\dagger|\hat{\bf p}_1,S,m_S\rangle_I{}_I\langle \hat{\bf p}_1,S,m_S|T|i\rangle
  (2\pi)^4\delta^4(p_1+p_2-p)\,.
\end{equation}
\end{widetext}
Inserting the second equation in Eq.~(\ref{eq:X_momentum2}) and its analogue for
$_I\langle \hat{\bf p}_1,S,m_S|T|i\rangle$ and performing the momentum
integration in Eq.~(\ref{eq:unitarity3}) the orbital part drops out
due to the orthogonality of the spherical harmonic functions and the
spin dependent part simply gives $\sum_m\left|\left\langle
    J,M|L,m,S,M-m\right\rangle\right|^2=1$ due to the unitarity of the
Clebsch-Gordan coefficients. One ends up with the reduced velocity
\begin{equation}\label{eq:beta_eff}
\beta_I\equiv \frac{2\beta_1\beta_2}{\beta_1+\beta_2}\,,
\end{equation}
where 
$\beta_i\equiv|{\bf p}_i|/E_i$ are the
velocities of the two intermediate state particles. In the center of mass frame, and
for equal mass particles, this reduces to $\beta_I=\beta_1=\beta_2=\sqrt{1-4m_I^2/s}$. 
Overall, one gets
\begin{equation}\label{eq:unitarity4}
  \Im\left[{\cal M}_{i\to f}(s)\right]=
   \sum_I \frac{\beta_I}{2^6\pi^2}\sum_{L,S} {\cal M}_{i\to (I,L,S)}(s)
    {\cal M}^*_{f\to (I,L,S)}(s)\,,
\end{equation}
where the sum over $L$ and $S$ runs over all combinations which can
combine to $J$ and are allowed by possible other conserved
quantum numbers carried by the intermediate state, such as parity. 

As schematically shown in Fig.~\ref{fig:OT}, Eq.~(\ref{eq:unitarity3}) 
relates two tree level amplitudes to a 1-loop process
by cutting the loop along the dashed lines and putting the intermediate states on-shell. Note
that for the special case $|i\rangle=|f\rangle$ the right hand side of
Eq.~(\ref{eq:unitarity3}) is essentially the total cross section
$\sum_I\sigma_{i\to I}$ into all possible intermediate states which is
thus related to the imaginary part of the
 forward scattering amplitude $\left\langle
  i|T|i\right\rangle={\cal M}_{i\to i}(p)$. This is known as the
{\it optical theorem}.

The transition rates $\Gamma_{i\to I}$ from a given initial state
$|i\rangle$ to any state $I$ consisting of two particles with not
necessarily equal masses are proportional to the phase space integral on the
right hand side of Eq.~(\ref{eq:unitarity3}) taken for
$|f\rangle=|i\rangle$. With Eq.~(\ref{eq:unitarity4}), and including the appropriate normalization factors, this reads

\begin{equation}\label{eq:rates}
  \Gamma_{i\to I}(s)\propto v\sigma_{i\to I}(s)=\frac{\beta_I}{2^5\pi^2 s}\sum_{L,S}|{\cal M}_{i\to (I,L,S)}(s)|^2\,.
\end{equation}
While we keep here all expressions fully general, we will later mostly consider the $v\equiv2\beta_i\rightarrow0$ limit and thus restrict ourself to $s$-wave annihilation, motivated by the fact that $p$-wave contributions are strongly suppressed with a factor of $v^2\sim10^{-6}$ for  typical galactic DM velocities  and therefore could  in any case not account for the observed line signal.
Squaring Eq.~(\ref{eq:unitarity4}), using the Cauchy-Schwarz
inequality $\left|\sum_i
  a_ib_i\right|^2\leq\left(\sum_i|a_i|^2\right)\left(\sum_i|b_i|^2\right)$
and combining with Eq.~(\ref{eq:rates}) yields

\begin{widetext}
\begin{equation}\label{eq:ratios}
  \frac{ r_{i\to f}\sigma_{i\to f}}{\sum_I\sigma_{i\to
      I}}=\frac{ r_{i\to f}\Gamma_{f}}{\sum_I\Gamma_{i\to I}}\leq
  \frac{N_f\beta_f}{2^{12}\pi^4}\sum_{I,L,S}\beta_I|{\cal
    M}_{(I,L,S)\to f}(s)|^2= \frac{N_f\beta_f^2s}{2^6\pi^2}\sum_I\sigma_{f\to I}\,,
\end{equation}
\end{widetext}
where
\begin{equation}
  \label{rdef}
  r_{i\to f}\equiv \frac{\left(\Im\left[{\cal M}_{i\to f}\right]\right)^2}
  {\left|{\cal M}_{i\to f}\right|^2}\leq1\,.
\end{equation}
The symmetry factor $N_f$ introduced above is $1/2$ for two photon final states and $1$ otherwise; note that such a factor does not appear in Eq.~(\ref{eq:rates}) because we are only interested in charged intermediate particles.

If there is only one intermediate state $(I,L,S)$ or if one of them
dominates, Eq.~(\ref{eq:ratios}) turns into an
equality. We note that if the absolute value of the amplitudes ${\cal M}_{i\to (I,L,S)}$ and $ {\cal M}^*_{f\to (I,L,S)}$ do not
correlate, the limit Eq.~(\ref{eq:ratios}) is very conservative and one should
resort to the original sum in Eq.~(\ref{eq:unitarity4}) to derive limits.
In order to proceed in that case, one  has to specify an (effective) Lagrangian 
 which describes the interaction between $\chi$ and $I$, see Appendix \ref{app:OT2}.

Eq.~(\ref{eq:ratios}) is our master equation. We will apply it to the
case where the final state $|f\rangle$ is $\gamma\gamma$, $\gamma Z$
or $\gamma h$, 
summed over all polarization states with the same $J^P$ value as the 
initial state $i$ of the two annihilating DM particles,
and the intermediate states $|I\rangle$ consist of a particle-anti-particle
pair. Since the angular momentum orientation independent part ${\cal
  M}_{(I,L,S)\to f}$ of the tree level amplitude defined in
Eq.~(\ref{eq:X_momentum2}) can straightforwardly be
calculated from the Standard Model or its extensions and the tree level
cross section $\sigma_{i\to I}$ is constrained from observations to be
discussed in Sect.~\ref{sec:constraints}, we will get constraints on
$ r_{i\to f}$ times the 1-loop cross section $\sigma_{i\to f}$  giving rise to
gamma-ray lines; for this quantity, we will sometimes use the short-hand notation.
\be
\label{imsv}
\Im (\sigma v)_f\equiv r_{i\to f} (\sigma v)_{i\to f}\,,
\ee
where the right-hand side is understood to be summed over all polarization states as explained above.

\begin{table*}[th!]
 \centering
 \begin{tabular}{c||c|c|c|c}
       {\bf Majorana} & 'final' state $I$ & \multirow{2}{*}{$ \frac{\Im (\sigma v)_{\gamma\gamma}}{(\sigma v)_{\rm tree}}$}  &  \multirow{2}{*}{$\frac{\Im(\sigma v)_{\gamma Z}}{(\sigma v)_{\rm tree}}$}  &  \multirow{2}{*}{$\frac{\Im(\sigma v)_{\gamma H}}{(\sigma v)_{\rm tree}}$} \\
       {\bf  DM} & $(L,S)$ &  &   \\
\hline\hline
  $\chi\chi\rightarrow \bar f f$ & $(0,0)$ & $\frac{N_cQ^4\alpha_{\rm em}^2m_f^2}{2m_\chi^2}\frac{1}{\beta_I}\left[\tanh^{-1}\beta_I\right]^2$ & $\frac{N_cQ^2\alpha_{\rm em}\alpha m_f^2}{\cos^2\theta_W m_\chi^2}\left[\frac{T_3}{2} - Q \sin^2\theta_W \right]^2\frac{ \beta_f}{\beta_I}\left[\tanh^{-1}\beta_I\right]^2$ & 0  \\
\hline  
  $\chi\chi\rightarrow W^+W^-$  & $(1,1)$ & $2 \alpha_{\rm em}^2 \beta_I\left[\tanh^{-1}\beta_I\right]^2$ & $4 \alpha_{\rm em}\alpha \cos^2\theta_W \beta_f\beta_I\left[\tanh^{-1}\beta_I\right]^2$ & 0  \\
  \end{tabular}
 \caption{\label{tab:OT_Majorana}  For Majorana DM, the initial state has $J=0^-$ in the zero-velocity limit, leaving only the above stated spin and orbital angular momentum combinations for charged 'final' state particles $I$. The quoted relations between tree-level and loop-level annihilation rates,
 with $\beta_f=1-m^2_Z/4m^2_\chi$ and $\beta_I^2=1-m_{f(W)}^2/m_\chi^2$, then follow as described in the text, c.f.~Eqs.~(\ref{eq:ratios}--\ref{imsv}), and  Appendix \ref{app:OT}.}
\end{table*}

\begin{table*}[th!]
 \centering

 \begin{tabular}{c||c|c|c|c}
       {\bf Scalar} &'final' state $I$  & \multirow{2}{*}{$ \frac{\Im (\sigma v)_{\gamma\gamma}}{(\sigma v)_{\rm tree}}$}  &  \multirow{2}{*}{$\frac{\Im(\sigma v)_{\gamma Z}}{(\sigma v)_{\rm tree}}$}  &  \multirow{2}{*}{$\frac{\Im(\sigma v)_{\gamma H}}{(\sigma v)_{\rm tree}}$} \\
       {\bf  DM} & $(L,S)$ & &     \\
\hline\hline
  $\chi\chi\rightarrow \bar f f$ & (1,1) & $\frac{N_cQ^4\alpha_{\rm em}^2m_f^2}{2m_\chi^2}\beta_I[\tanh^{-1}\beta_I]^2$ & $\frac{N_cQ^2\alpha_{\rm em}\alpha m_f^2}{\cos^2\theta_W\beta_f\beta_I m_\chi^2}\left[\frac{T_3}{2} - Q \sin^2\theta_W \right]^2$\qquad\qquad\qquad\qquad & 0    \\
{} & {} & {} & \qquad $\times\left[\beta_I\tanh^{-1}\beta_I - (1-\beta_f)\left(\frac{\tanh^{-1}\beta_I}{\beta_I}-1\right)\right]^2$ &   \\
\hline 
  $\chi\chi\rightarrow W^+W^-$  & $\ket t\!\!\equiv\!\!\sqrt{\frac23}\ket{0,0}$ & 
  $\alpha_{\rm em}^2\frac{\left(1+\beta_I^2\right)^2}{2\beta_I}[\tanh^{-1}\beta_I]^2$ & 
  $\frac{\alpha_{\rm em}\alpha \cos^2\theta_W}{\beta_I\beta_f}\Big{[}(\beta_f^2+\beta_I^2)\tanh^{-1}\beta_I$ \qquad\qquad\qquad& 0   \\
{} & $\phantom{\ket t\!\!=\!\!}+\!\frac1{\sqrt 3}\ket{2,2}$ & {} & 
\qquad\qquad\qquad$+\frac{(\beta_f-\beta_I^2)}{\beta_I}(1-\beta_f)\left(\frac{\tanh^{-1}\!\!\beta_I}{\beta_I}\!-\!1\right)\Big{]}^2\!\!$ & {}  \\
{} & 
$\ket l\!\!\equiv\!\!\sqrt{\frac23}\ket{2,2}$ & $\alpha_{\rm em}^2\frac{\left(1-\beta_I^2\right)^2}{4\beta_I}[\tanh^{-1}\beta_I]^2$ & 
$\frac{\alpha_{\rm em}\alpha\cos^2\theta_W}{2\beta_I\beta_f}\frac{m_W^4}{m_\chi^4}\Big{[}\left(1\!-\!\frac{m_Z^2}{4m_W^2}\!-\!\frac{m_Z^4\beta_I^2}{8m_W^4}\right)\tanh^{-1}\!\beta_I$ & 0  \\
{} & 
$\phantom{\ket l\!\!=\!\!}-\!\frac1{\sqrt 3}\ket{0,0}$ & {} & 
\qquad$-\frac{m_Z^2}{4\beta_I m_W^2}\left(2\!-\!\beta_I^2\!-\!\frac{m_Z^2}{2m_W^2}\right)\left(\frac{\tanh^{-1}\!\!\beta_I}{\beta_I}\!-\!1\right)\Big{]}^2$ & {} \\
\end{tabular}
 \caption{\label{tab:OT_Scalar}  Same as Tab.~\ref{tab:OT_Majorana}, but for Scalar DM (where the initial state has $J=0^+$ in the zero-velocity limit). Note that in this case there are two independent degrees of freedom for the $W$-boson 'final' states $I$; here, we state the results for longitudinal and transverse states (see also Appendix \ref{app:OT1} for more details). }
\end{table*}

For a scalar initial state, $J=0$, the sum reduces to a sum over $L=S$ only; this is
for example the case for {\it Majorana DM} in the $v\rightarrow0$ limit we are interested 
in here. In this case the initial state has also negative parity, implying that 
fermionic (bosonic) states $I$ must combine to $L=S=2n$ ($L=S=2n+1$), with 
$n=0,1,2...$. Table~\ref{tab:OT_Majorana} shows the corresponding ratios 
of Eq.~(\ref{eq:ratios}) for all relevant channels.
The results for {\it scalar DM}, where the initial state has positive parity, are given
in Tab.~\ref{tab:OT_Scalar}; here, $L=S=2n~(2n+1)$ for bosonic (fermionic) states $I$ -- 
which implies that there are {\it two} independent degrees of freedom for $W$ boson loops (see also the discussion in Appendix \ref{app:OT1}). 
In the following, we will limit our discussion mostly to the Majorana case but note that an 
extension of our analysis to DM particles with any spin properties in the initial
state is straight-forward with the formalism presented here.  
For more details about the calculation leading to the results presented in 
Table~\ref{tab:OT_Majorana}, we refer the reader to Appendix \ref{app:OT}. 
We checked explicitly that the two methods described there give consistent results, 
and that  the results for $\gamma\gamma$ final states agree with those reported in 
Ref.~\cite{Abazajian:2011tk}.

Let us finally point out that in the case of $\gamma Z$ and $\gamma h$ final states dominated by fermion loops, there is a contribution to the imaginary part of the cross section, at the same order in the coupling constants, that is not
captured in Eq.~(\ref{eq:ratios}) above. This is related to the fact that both $Z$ and $h$ are unstable, such that there is another way of cutting through the loop diagram shown in Fig.~\ref{fig:OT} (as indicated by the dotted line). While the possibility of DM
annihilation to such three-body ($\bar f f\gamma$) final states is obviously not taken into account in our formalism leading to Eq.~(\ref{eq:ratios}), we note that those processes are $\mathcal{O}(\alpha_{\rm em})$-suppressed with respect to the tree-level annihilation (to $\bar f f$) and therefore  only marginally affect the general limits on the annihilation rate that we will discuss in the next Section. While the contribution to the imaginary 
part of the loop amplitude still may be sizable, explicit calculation for Majorana DM along the lines explained in Appendix  \ref{app:OT2} shows that those contributions always come with opposite signs as compared to the contributions from Eq.~(\ref{eq:ratios}). Neglecting them will thus lead to an over-estimation of the imaginary part and thus {\it conservative} bounds on the quantity defined in Eq.~(\ref{imsv}).

\section{Constraints on the annihilation rate}
\label{sec:constraints}

At tree-level, DM annihilates or decays into SM particles which then (except for electrons and neutrinos) fragment and/or decay into stable particles like photons (mostly via $\pi^0\rightarrow\gamma\gamma$), antiprotons and electrons or positrons. In this Section, we will discuss the various constraints on the decay or annihilation rate that derive from the non-observation of excesses in gamma-rays, cosmic ray antiprotons and radio data. We will focus on DM masses that could explain the observed line feature at  $E_\gamma^{130}=129.8^{+7.4}_{-13.2}$\,GeV \cite{Weniger:2012tx},
which in the case of annihilating DM implies
\be
\label{mchi}
m_\chi = \frac{E_\gamma^{130}}{2}\left[1+\sqrt{1+\left(m_X/{E_\gamma^{130}}\right)^2}\right]\,,
\ee
where $X=\gamma,Z,h$ denote the most interesting possibilities (though in principle $X$ may also be a new neutral state \cite{Bertone:2009cb}). Decaying DM would require $m_\chi=E_\gamma^{130}+\sqrt{\left(E_\gamma^{130}\right)^2+2m_X^2}$, but we will not consider this option in the following because it is not supported by the observed angular distribution of the signal \cite{line_review,Buchmuller:2012rc}.

For all of this section, we will for consistency use an Einasto profile for the DM distribution in the Milky Way,
\be
\label{einasto}
\rho_\chi(r)=\rho_0\exp\left(-\frac{2}{\alpha}\frac{r^\alpha}{r_s^\alpha}\right),
\ee
with parameters chosen such as in Ref.~\cite{Weniger:2012tx}, i.e.~$\alpha=0.17$, $r_s=20\,$kpc and  $\rho_0=1.05\cdot10^4\,{\rm GeV}\,{\rm cm}^3$ (which results in $\rho_\chi(R_\odot)=0.4\,{\rm GeV}\,{\rm cm}^3$ at the position of the sun, $R_\odot=8.5\,{\rm kpc}$). Such a profile is not only favored by $N$-body simulations of gravitational clustering \cite{Springel:2008cc} but also provides a very nice fit to the line observation \cite{line_review}.

\subsection{Continuum gamma rays}

Dwarf spheroidal galaxies exhibit mass-to-light ratios of up to $\mathcal{O}(1000)$, which are the largest values observed in astronomical objects. If located at not too large distances from the sun, this makes them ideal targets for DM detection because the gamma-ray emission expected from astrophysical processes is negligible \cite{Tyler:2002ux,Evans:2003sc}. In fact,
 the non-observation of any gamma-ray signal from dwarf galaxy satellites of the Milky Way by Fermi places the currently strongest constraints on secondary photons from DM annihilation \cite{Ackermann:2011wa}. We collect these constraints for all relevant  channels in Tab.~\ref{tab:constraints}, conservatively choosing the weakest bound in the ranges for the DM masses of interest, as defined by Eq.~(\ref{mchi}). We note that constraints on the $e^\pm$ channel are not explicitly provided in Ref.~\cite{Ackermann:2011wa}. However, the photon spectrum from $e^\pm$ is very similar to the one from $\mu^\pm$, which means that a good estimate for bounds on $e^\pm$ final state is obtained by rescaling the $\mu^\pm$ bounds by the number of photons produced per annhilation, $N_\gamma=\int_{1\,{\rm GeV}}^{m_\chi} dE_\gamma\,dN_\gamma/dE_\gamma$, which in both cases is given by final state radiation \cite{Bergstrom:2004cy}.
 
The Fermi dwarf limits assume a DM density profile that scales as $r^{-1}$ in the innermost part.
Such an NFW profile is also consistent with numerical  $N$-body simulations \cite{Navarro:1995iw} and only slightly steeper in the central part than our reference profile of Eq.~(\ref{einasto}). While kinematical data presently cannot distinguish between  shallow or (possibly very) cuspy central profiles in dwarf galaxies, the resulting limits on the DM annihilation rate can change sizably. For that reason, a more conservative approach relies on considering instead limits on the annihilation rate that derive from continuum gamma rays in the galactic center region itself; since we will eventually be interested in the {\it ratio} of photon fluxes for the continuum and line component, this also minimizes the astrophysical uncertainties related to the DM density profile. 

In Tab.~\ref{tab:constraints} we therefore also present a compilation of the corresponding limits from a recent analysis of a region with $|b|<5^\circ$ and $|l|<5^\circ$ around the galactic center \cite{Cholis:2012fb}.
Those limits were derived from a slightly different DM profile than in Eq.~(\ref{einasto}); 
we therefore divided them by a factor of 1.4, taking into account that they should roughly scale with the quantity $J\equiv\int_{\Delta\Omega} d\Omega\int_{l.o.s.}\!\!\!\!\!ds\, \rho^2(r)$ (where $\Delta\Omega$ corresponds to the angular region considered and $l.o.s$ refers to a line-of-sight integration).
 Note that the limits on light lepton final states are actually {\it stronger} than for the dwarf analysis  because of inverse Compton scattering of high-energy $e^\pm$, which upscatters low-energy photons of the rather dense interstellar radiation field to gamma-ray energies.

\begin{table*}[th!]
 \centering
\begin{tabular}{c||c|c|c|c|c|c}
{\bf general} & cont. gamma & cont. gamma & antiprotons & antiprotons &
synchrotron & synchrotron \\
{\bf WIMP} & (dwarfs) & (GC) & ('KRA', $L=4$\,kpc) & ('CON',
$L=10$\,kpc) & (full cone) & ($r<1\,$pc) \\
\hline\hline
$b\bar b$ & 7.6 (8.1, 8.6) & 21 (22, 23) & 10.4 (11.6, 11.5) & 4.2 (4.7,
4.7) & 27.5 (29.6, 31.2) & 89 (101, 110) \\
\hline
$\tau^+\tau^-$ & 16 (18, 20) & 14 (15, 16) & --- & ---& 25.8 (29.6,
32.7) & 369 (441, 500) \\
\hline
$\mu^+\mu^-$ & 145 (168, 190) & 28 (28, 29) & --- & ---& 18.2 (21.8,
24.7) & 427 (515, 589)\\
\hline
$e^+e^-$ & 89 (104, 118) & 14 (11, 13) & --- & --- & 16.1 (19.4, 22.2) &
419 (506, 579)\\
\hline
$W^+W^-$ & 11 (12, 12) & 24 (24, 26) & 9.3 (9.5, 9.8) & 3.8 (3.9, 4.0) &
29.7 (32.5, 34.7) & 122 (139, 152) \\
\end{tabular}
 \caption{\label{tab:constraints}  This table shows the maximally allowed DM annihilation rate (in units of $10^{-26}$cm$^3$s$^{-1}$) into the stated channels, assuming a DM mass consistent with a $\gamma\gamma$ ($\gamma Z, \gamma h$) interpretation of the line observation. See text for details about  the various limits and underlying assumptions.}
\end{table*}

\subsection{Antiprotons}

The propagation of antiprotons can very well be described in simple phenomenological diffusion models, the parameters of which are strongly constrained by other cosmic ray data like in particular the boron over carbon ratio $B/C$  \cite{Donato:2003xg}. In this way one can predict the astrophysical background (dominated by secondary antiprotons from cosmic ray proton collisions with interstellar hydrogen) with remarkably small uncertainties. The flux of primary antiprotons from DM annihilation, on the other hand,  is subject to greater uncertainties; these are mostly related to the vertical size $L$ of the magnetic halo which is directly proportional to the annihilation volume being probed. While the $B/C$ analysis leaves a degeneracy between the diffusion strength and $L$, nominally allowing values as small as $L\sim1\,$kpc, considerably larger values ($\sim$4--10\,kpc) are preferred when also taking into account radioactive isotopes \cite{Putze:2010zn}, gamma rays~\cite{Timur:2011vv}, cosmic-ray electrons~\cite{electronL} or radio data \cite{Bringmann:2011py,DiBernardo:2012zu}. 

We will therefore mainly use the reference ('KRA') model for cosmic ray propagation from a recent  comprehensive study \cite{Evoli:2011id}, featuring $L=4\,$kpc. As the most optimistic model in terms of constraining DM annihilation, we will also refer to the 'CON' model of the same analysis (with $L=10\,$kpc). We use  \ds~\cite{DS} to calculate the propagation of primary antiprotons and the expected flux at the top of the atmosphere (building on the procedure described in Refs.~\cite{Evoli:2011id, Bergstrom:1999jc}). 
In order to derive limits on the annihilation cross-section, we then demand that the 
minimally expected astrophysical background of secondary
antiprotons  (which we take from Ref.~\cite{Bringmann:2006im}) plus the 
DM signal  do not overshoot the antiproton measurements by
PAMELA~\cite{Adriani:2010rc} by more than $3\sigma$ in any data point. For our purpose,
the only relevant data points turn out to be those between 7 and 26.2 GeV; 
solar modulation and convection affect the antiproton flux only at energies smaller than about 1 GeV and the associated uncertainties therefore have no impact on our limits.

For comparison, we show our results in  Tab.~\ref{tab:constraints} for the same annihilation channels as for the gamma-ray constraints. 
In principle, final state radiation of electroweak gauge bosons in the case of  lepton final states would also produce antiprotons \cite{Ciafaloni:2010ti}  -- albeit at rather low rates for the small values of $m_\chi$ we are interested in here. 
We thus do not include those final states because the resulting constraints would anyway be much weaker than from gamma rays or synchrotron radiation (see below).


\subsection{Synchrotron radiation}

Let us now turn to the synchrotron radiation emitted by electrons from
WIMP annihilation at or near the galactic center.  We use a semi-analytical approach to derive  
constraints from radio observations, in particular the $\SI{50}{mJy}$ upper limits on the 
flux density at $408\,$MHz \cite{Davis:1976}, and summarize 
our results in Table \ref{tab:constraints}.

We first note that in a magnetic field of strength $B$ the peak
contribution to synchrotron emission at a given frequency $\nu$ comes
from electrons and positrons with an energy roughly twice the critical energy,
\bea
\label{eq:E_crit}
  E_p(\nu)\simeq0.29^{-\frac12}E_c&\equiv&
 \left(\frac{4 \pi m_e^3 \nu}{3\cdot 0.29\,e B}\right)^\frac12\\
  &=&0.46\left(\frac{\nu}{{\rm GHz}}\right)^\frac12\left(\frac{B}{{\rm mG}}\right)^{-\frac12}
  \,{\rm GeV}\,.\nonumber
\eea

As discussed in Ref.~\cite{Bertone:2001jv}, close to the galactic center
the electrons and positrons are deeply in the diffusive regime on time
scales over which they lose most of their energy due to synchrotron
radiation. This becomes clear in quantitative terms when comparing
the energy loss time for synchrotron emission $t_{\rm loss}$ and the diffusion
coefficient $D$ evaluated at the electron energy $E_p(\nu)$,
\begin{align}\label{eq:lenghts}
t_\mathrm{loss}\simeq&\frac34\left(\frac{m_e^5}{\pi e^7}\right)^{1/2}\frac1{\nu^{1/2}}\frac1{B^{3/2}}\ ,\\
D&\simeq\frac23\left(\frac{\pi m_e^3}{e^3}\right)^{1/2}\frac{\nu^{1/2}}{B^{3/2}}\,.
\end{align}
Here, as usual, the diffusion coefficient is one third of the
effective scattering length on the magnetic inhomogeneities which was
approximated by the gyration radius, corresponding to the Bohm limit. Since
$D/t_\mathrm{loss}\sim 
r_e\nu\simeq4\times10^{-15}\,(\nu/408\,{\rm MHz})$, where
$r_e=e^2/m_e$ is the classical electron radius, we can use the
diffusion approximation to estimate the length scale $l_\mathrm{diff}$ over which
electrons propagate during their energy loss time. This gives
$l_\mathrm{diff}\simeq(D\,t_\mathrm{loss})^{1/2}$,
which explicitly depends on the magnetic field profile. A fairly well motivated, though somewhat simplistic, model for the value of $B$ close to the central black hole (BH) can be constructed by assuming steady accretion and thus equating magnetic and matter kinetic energies \cite{1992ApJ387L25M}, leading to a $B\sim r^{-5/4}$ behavior:
\begin{equation}\label{eq:magfield}
 B(r)=\SI{7.2}{mG}\times\begin{cases}
          (R_{\textrm{acc}}/r)^{5/4} & r<R_{\textrm{acc}}\\
          (R_{\textrm{acc}}/r)^{2} & R_{\textrm{acc}}<r\lesssim100R_\textrm{acc}\\
          10^{-4} & r\gtrsim 100R_\textrm{acc}
         \end{cases}\ ,
\end{equation}
where $R_\textrm{acc}\approx\SI{.04}{pc}$ corresponds to the accretion radius of the BH.

The second radial range of Eq.~(\ref{eq:magfield}), where $B\propto r^{-2}$ follows from the assumption of magnetic flux conservation, turns out to be the
most relevant for the synchrotron emission considered here. Noting that in the
vicinity of the galactic center, the dark matter density scale height $l_\chi\equiv
|\rho_\chi(r)/\rho^\prime_\chi(r)|$ equals
$(r_s/2)(r/r_s)^{1-\alpha}$ for the Einasto profile (\ref{einasto}), the ratio
$l_\mathrm{diff}/l_\chi$  becomes
\be
\frac{l_\mathrm{diff}}{l_\chi}\simeq\frac{m_e^2c^4}{\sqrt2e^{5/2}l_\chi B^{3/2}}\simeq\num{.3}\left(\frac{r}{{\rm pc}}\right)^{2+\alpha}\,.
\ee
 For distances $r\lesssim\SI{1}{pc}$, electrons and positrons
thus basically loose their energy {\it in situ} which implies that the
local synchrotron emission power is just proportional to the local
annihilation rate into electron-positron pairs. In contrast, at radii
$r\gg\SI{1}{pc}$ diffusion washes out the radio emission profile
compared to the local pair injection distribution. 

Carrying out the same procedure as in
\cite{Bertone:2001jv,Bertone:2008xr}, where 
diffusion is neglected and the
monochromatic approximation on single-electron synchrotron spectra
Eq.~(\ref{eq:E_crit}) is
used, we then obtain the following formula for the total synchrotron flux density
 for a Majorana WIMP (an additional factor of 1/2 is necessary if
the WIMP is a Dirac fermion):
\begin{equation}\label{eq:synpower}
F_\nu\equiv\frac1{4\pi R_\odot^2}\frac{\dif W_{\textrm{syn.}}}{\dif\nu}\approx\frac{(\sigma v)}{8\pi\nu R_\odot^2 M_\chi^2}\int\!\!E_p\rho_\chi^2(r)N_e(E_p)\dif V\ , 
\end{equation}
where we consider two different integration regions: the first  is a cone with half-aperture of $\SI{4}{\arcsec}$, corresponding to the $\SI{408}{MHz}$ observation \cite{Davis:1976}, while the second is defined as the intersection of this cone with a sphere of radius $\SI{1}{pc}$ around the GC (which roughly sets the limit where Eq.~(\ref{eq:synpower}) is strictly valid).  $N_e(E_p)$ is the number of $e^\pm$ pairs above the peak energy  $E_p$ that is produced per annihilation and is obtained numerically from  \ds\ for the various annihilation channels considered here.

In Table \ref{tab:constraints}, we report the corresponding bounds on $\sigma v$ that result from the observed 50\,mJy upper  limit on the flux at $\SI{408}{MHz}$. We note that the bound obtained for the truncated integration region is clearly very conservative since there are also contributions to the total synchrotron flux in the line of sight that come  from radii larger than the cutoff distance $\SI{1}{pc}$. These contributions, however, require in principle the full solution of the diffusion equation and are therefore notably harder to compute. Simply integrating Eq.~(\ref{eq:synpower}) over the full observation volume, on the other hand, results in somewhat optimistic constraints as it overestimates the actual  density of the electrons that produce synchrotron radiation. Nevertheless, this is the typical approach taken (see  e.g.~Refs.~\cite{Bertone:2008xr,Laha:2012fg}) and we will adopt it in the following in order to allow for a simple comparison.

Let us finally stress that Eq.~(\ref{eq:synpower}) is fortunately only mildly sensitive to the in principle rather uncertain central magnetic field distribution. The radio constraints resulting from an alternative (weak) B-field model assumed to be constant for $r<R_\textrm{acc}$, e.g., differ by just $\sim1\%$. 
The dependence of our radio constraints on the adopted innermost behavior of the DM profile, on the other hand, is much stronger. The most extreme case corresponds to assuming a {\it constant} DM density below an angular distance of 1$^\circ$ around the GC, in which case our constraints weaken by around two orders of magnitude. We note, however, that the observed signal is consistent with an Einasto profile down to at least $\sim1^\circ$ \cite{line_review} and that there is no indication from $N$-body simulations in favor of such an abrupt change of the profile at this scale. 

\begin{table*}[th!]
 \centering
 \begin{tabular}{c||c|c|c||c}
       {\bf Majorana} & cont. gamma {\it limit} &antiproton {\it limit} &  synchrotron {\it limit} & {\it expected} value\\ 
       {\bf WIMP} &  (GC) & ('KRA', $L=4$\,kpc) & (full cone)  
       & (effective operator) \\ \hline\hline
 $b\bar b$ & $6.0\times10^{-6}$ ($5.3\times10^{-6}$) & $3.0\times10^{-6}$  ($2.8\times10^{-6}$)  &
 $7.8\times10^{-6}$ ($7.2\times10^{-6}$)  &$0.44$ ($0.13$)   \\
\hline   
 $\tau^+\tau^-$ & $2.9\times10^{-5}$ ($3.5\times10^{-8}$) & --- & 
 $5.3\times10^{-5}$ ($6.8\times10^{-8}$)  & $0.33$ ($0.089$) \\
\hline   
 $\mu^+\mu^-$   & $5.1\times10^{-7}$ ($5.6\times10^{-10}$)  & --- & 
 $3.3\times10^{-7}$ ($4.4\times10^{-10}$) & $0.15$ ($0.036$) \\
\hline   
 $e^+e^-$  & $1.7\times10^{-11}$ ($1.4\times10^{-14}$) & --- & 
 $1.9\times10^{-11}$ ($2.6\times10^{-14}$) & $0.055$ ($0.013$)\\
\hline   
  $W^+W^-$    & 0.021 (0.074) & $7.9\times10^{-3}$ (0.029) & 
  0.025 (0.10) & $0.048$ ($0.092$)
  \end{tabular}
 \caption{\label{tab:limits_Majorana}  This table shows the maximally allowed relative contribution $r$ of the imaginary part of the amplitude to the total annihilation rate of Majorana DM particles into $\gamma\gamma$ ($\gamma Z$), as defined in Eq.~(\ref{rdef}), if the latter is to provide a viable explanation of the observed Fermi gamma-ray line at 130\,GeV. The imaginary part is assumed to be dominated by the indicated SM particles running in the loop. For comparison, the right column shows the {\it expected} value of $r$ when the gray blob in Fig.~\ref{fig:OT} is replaced by an effective  operator for a 3-point interaction between the respective pair of SM particles and a pseudoscalar that represents the annihilating DM particle pair; note that for the case of $\gamma Z$ final states dominated by fermion loops, this even includes the 'second cut' in Fig.~\ref{fig:OT} (which arises because the decay $Z\to\bar f f$ is possible and which is not included in our constraints -- see the discussion at the end of Section \ref{sec:OT}).}
\end{table*}

\begin{table*}[th!]
 \centering
 \begin{tabular}{c||c|c|c||c}
       {\bf Scalar} & cont. gamma {\it limit} &antiproton {\it limit} &  synchrotron {\it limit} & {\it expected} value\\ 
       {\bf WIMP} &  (GC) & ('KRA', $L=4$\,kpc) & (full cone)  
       & (effective operator) \\ \hline\hline
 $b\bar b$ & $6.0\times10^{-6}$ ($5.7\times10^{-6}$) & $3.0\times10^{-6}$  ($3.0\times10^{-6}$)  &
 $7.8\times10^{-6}$ ($7.7\times10^{-6}$)  &$0.45$ ($0.13$)   \\
\hline   
 $\tau^+\tau^-$ & $2.9\times10^{-5}$ ($3.7\times10^{-8}$) & --- & 
 $5.3\times10^{-5}$ ($7.2\times10^{-8}$)  & $0.33$ ($0.089$) \\
\hline   
 $\mu^+\mu^-$   & $5.1\times10^{-7}$ ($5.8\times10^{-10}$)  & --- & 
 $3.3\times10^{-7}$ ($4.5\times10^{-10}$) & $0.15$ ($0.036$) \\
\hline   
 $e^+e^-$  & $1.7\times10^{-11}$ ($1.5\times10^{-14}$) & --- & 
 $1.9\times10^{-11}$ ($2.6\times10^{-14}$) & $0.056$ ($0.012$)\\
\hline   
  $W^+W^-$ (t)   & 0.023 (0.076) & $8.8\times10^{-3}$ (0.030) & 
  0.028 (0.10) & \multirow{2}{*}{$0.094$ ($0.14$)} \\
  $W^+W^-$ (l)   & $1.2\times10^{-3}$ ($5.3\times10^{-4}$) & $4.5\times10^{-4}$ ($2.1\times10^{-4}$) & 
  $1.4\times10^{-3}$ ($7.1\times10^{-4}$) & 
  \end{tabular}
 \caption{\label{tab:limits_scalar}  Same as Tab. \ref{tab:limits_Majorana}, but for scalar DM (where the annihilating DM particle {\it pair} is also represented by a scalar).}
\end{table*}

\section{Implications for model building}
\label{sec:model}

For an Einasto profile as in Eq.~(\ref{einasto}), the annihilation rate to two photons that is required to fit the observed line signal is given by 
\be
\label{eq:signalnorm}
 \left\langle\sigma v\right\rangle_{\gamma\gamma}=1.27^{+0.37}_{-0.43}\times10^{-27}{\rm cm}^3{\rm s}^{-1}\,,
\ee
which translates into $\left\langle\sigma v\right\rangle_{\gamma Z}=3.14^{+0.89}_{-0.99}\times10^{-27}{\rm cm}^3{\rm s}^{-1}$  and $\left\langle\sigma v\right\rangle_{\gamma h}=3.63^{+1.02}_{-1.11}\times10^{-27}{\rm cm}^3{\rm s}^{-1}$ for the case of annihilation into $\gamma Z$  and $\gamma H$, respectively \cite{line_review}. Assuming that the line signal is caused by DM annihilation, we can now proceed to combine this information with the generic relation between tree- and loop-level annihilation rates 
discussed in Section \ref{sec:OT} and the constraints on the tree-level annihilation rate summarized in Table \ref{tab:constraints}.

For the case of Majorana DM (see Table \ref{tab:OT_Majorana}), the resulting constraints on how much the imaginary part of the amplitude may contribute to the loop process is shown in Table \ref{tab:limits_Majorana}; in order to be conservative, we always took the lower range of the allowed annihilation rate  in Eq.~(\ref{eq:signalnorm}) when deriving these values (as well as the lowest possible DM mass compatible with the signal, see Table 2 in Ref.~\cite{line_review}). As stressed before, these constraints assume that the loop signal is dominated by only one species of intermediate particles. In Table \ref{tab:limits_scalar}, we show for comparison the very similar constraints that arise for the case of scalar DM (see Table \ref{tab:OT_Scalar}). While beyond the scope of this work, we note that the computation of corresponding constraints for  DM particles with any other spin properties in the initial state is also straight-forward with the formalism presented here.

Obviously, fermionic loops are subject to the tightest bounds, especially for small fermion masses $m_f$. Such a behavior should be expected from  the helicity-suppressed annihilation of  $\gamma\gamma$ or $\gamma Z$ into a fermion pair,  which results in $\sigma_{f\rightarrow I}\propto m_f^2$ in Eq.~(\ref{eq:ratios}). We note that this scaling, taking into account also $r\propto Q^4$, can be used to derive very good approximations also to the limits on $r$  for other quark pairs $\bar q q$ than shown here; this is because the fragmentation functions to photons, antiprotons and electrons are very similar to the $\bar b b$ case, leading thus to essentially the same constraints on the tree-level annihilation rate. Charged gauge bosons do not exhibit such a suppression and are therefore not as strongly constrained; still, the imaginary part of a $W^\pm$ loop with  $\gamma\gamma$ ($\gamma Z$) final
states may at most contribute $\sim$1\% ($\sim$10\%) to the total rate.

Let us mention in passing that two-gluon final states provide very stringent complementary constraints if the loop is dominated by quarks \cite{Chu:2012qy}. For example, we expect 
$(\sigma v)_{gg}=2\alpha^2_{\rm s}/(9Q^4\alpha^2_{\rm em})(\sigma v)_{\gamma\gamma}\gtrsim 3\times10^{-26}Q^{-4}{\rm cm}^3{\rm s}^{-1}$
in case the observed line corresponds to a $\gamma\gamma$ final state \cite{Bergstrom:1988fp}. Using the fact that both antiproton and continuum
gamma-ray limits for $gg$ final states are very similar to the ones for $\bar b b$ 
listed in Table \ref{tab:constraints}, this implies that down-type quarks ($Q=-1/3$) in the loop are 
clearly excluded for a $\gamma\gamma$ signal while up-type quarks ($Q=2/3$) may still remain marginally consistent with those constraints. If the 130\,GeV feature is due to a $\gamma Z$ or $\gamma h$ loop, however, the corresponding annihilation rate into $gg$ depends on the loop topology and no model-independent conclusions are possible.

For comparison, we also calculated the full loop amplitude assuming that the 4-point interaction between DM and intermediate particles $I$ in Fig.~\ref{fig:OT} can be replaced  by an effective  operator for a 3-point interaction between $I$ and a (pseudo)scalar particle representing the initial DM pair (see also Appendix \ref{app:OT2}). This corresponds to the limit where all new physics states heavier than the DM particle are integrated out and therefore cannot contribute to the loop signal. If, as we always assume, only one type of SM particles contributes to the loop, this limit  does not change the result for the {\it imaginary} part of the amplitude; it may, however, in principle underestimate its {\it real} part. We report the numerical result of this calculation (using {\sf FeynCalc} \cite{Mertig:1990an} and {\sf LoopTools} \cite{Hahn:1998yk}) 
 in the last column of 
Table \ref{tab:limits_Majorana} and \ref{tab:limits_scalar}: as can be seen, it it always larger than the corresponding upper limits reported in the same Table. 

Even if also heavier particles contribute to the loop, which is the more commonly encountered situation in typical WIMP scenarios, the loop signal gets strongly enhanced if  the virtual particles can  (almost) be put on shell. The large annihilation rates necessary to explain the observed signal therefore seem to require rather generically that not only the real part, but also the imaginary part of the amplitude provides a significant contribution -- implying that {\it for large annihilation rates}, the values of $r$ in realistic models should not be too far below those obtained in the 
effective operator approach.
In that sense, our bounds present a serious challenge to any model-building attempt trying to attribute the observed 130\,GeV feature to a loop that involves SM particles, with the only possible exception of top quarks.\footnote{
Top loops obviously do not have an imaginary part because $m_t>m_\chi$. However, the contribution from top quarks to the loop is often connected to the contributions from other 
quarks in a simple and unique way (e.g.~through the Yukawa coupling with an $s$-channel Higgs); the imaginary part
of those loops may then be used to constrain even such scenarios.
}

\begin{figure*}[th!]
 \begin{center}
    \includegraphics[width=0.45\textwidth]{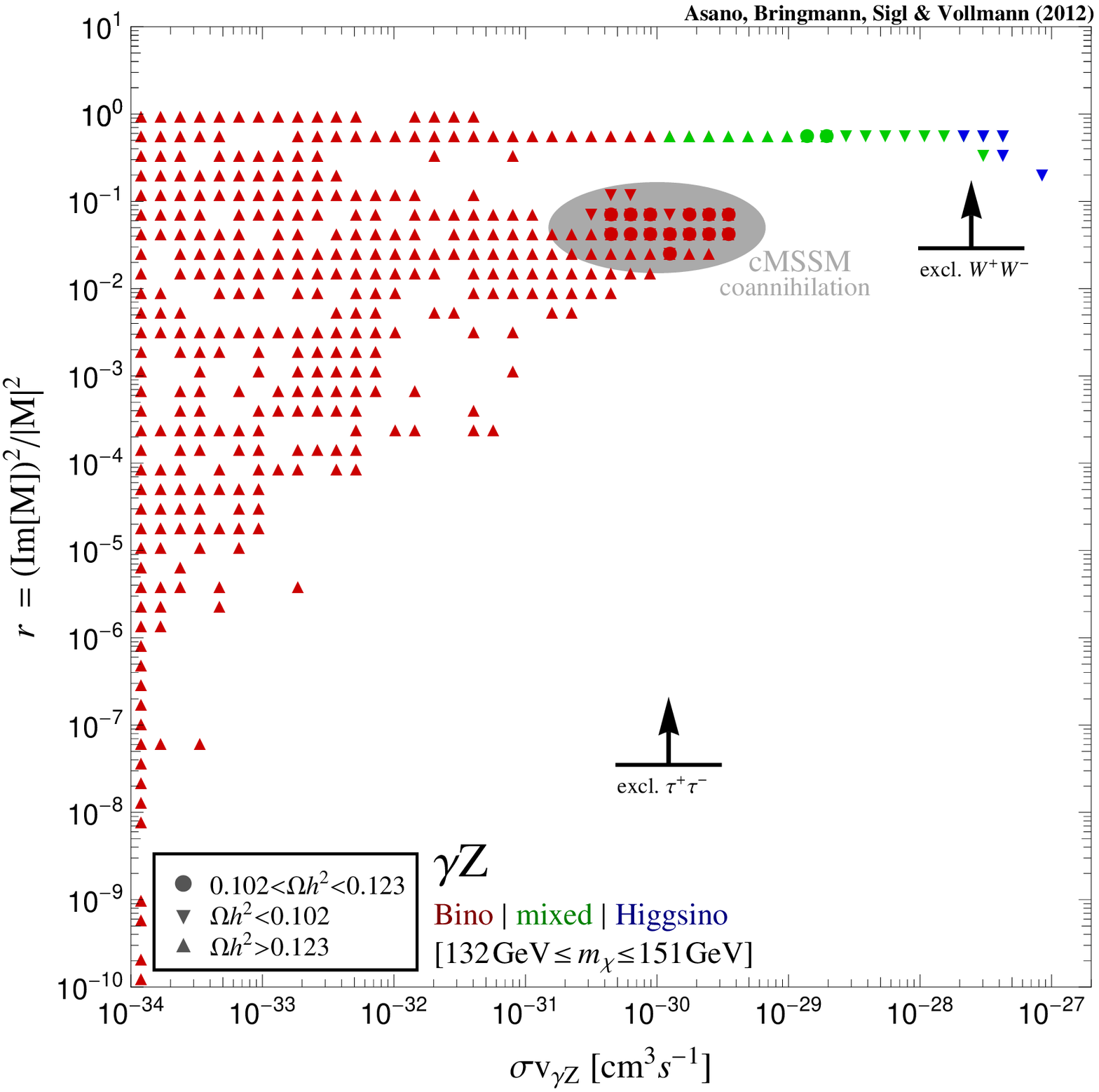}
    \includegraphics[width=0.45\textwidth]{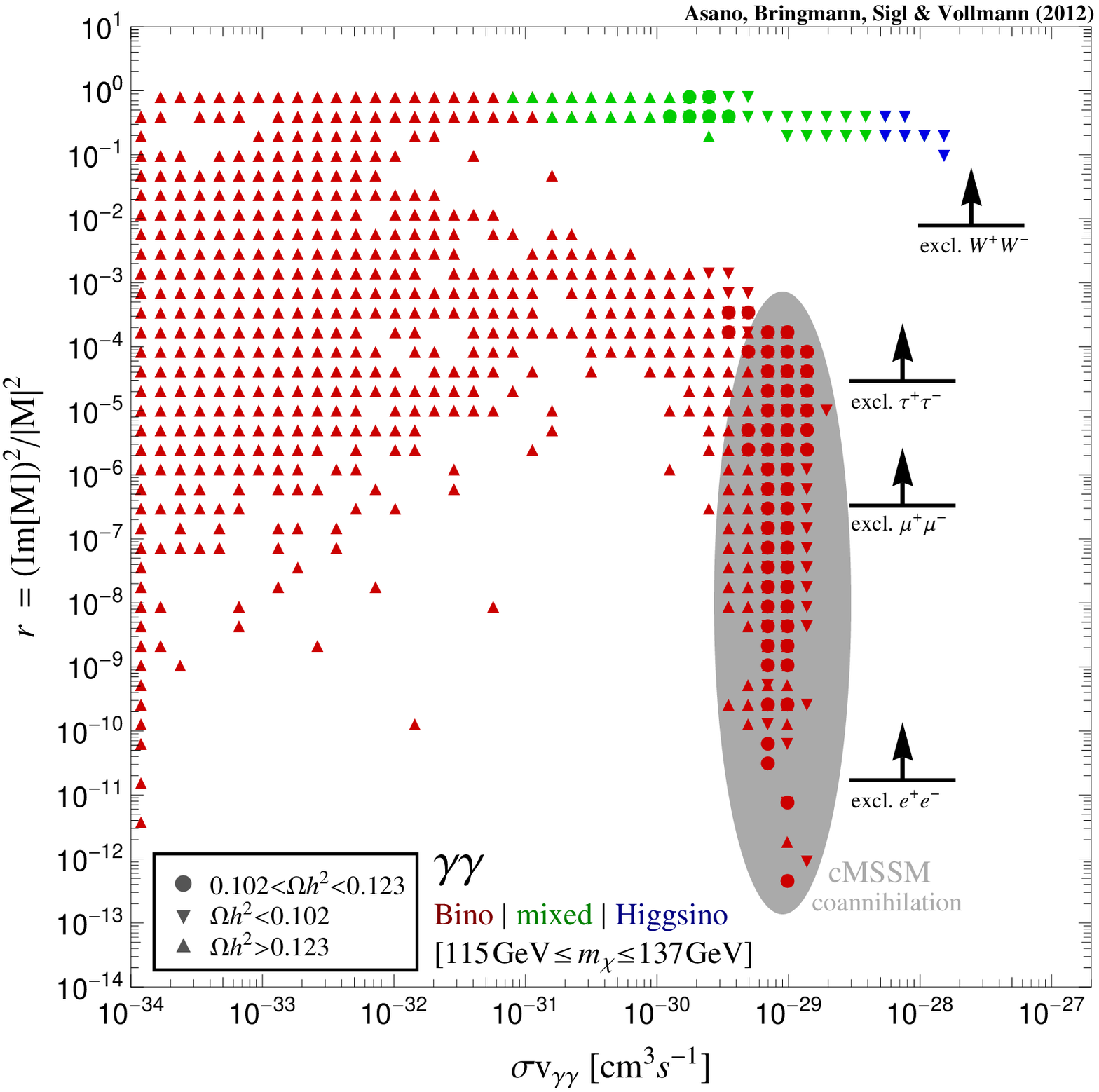}
 \end{center}
\caption{\label{fig:SUSYscan} Left: For a large scan over  MSSM and cMSSM models with $m_\chi\sim145\,{\rm GeV}$, this figure shows the loop-level annihilation rate $(\sigma v)_{\gamma Z}$ vs. the relative contribution $r$ of the imaginary part of the amplitude to this cross section. 
Models where the neutralino is dominantly a Higgsino (Bino) are indicated by blue (red) symbols; green symbols refer to mixed neutralino DM. Filled circles correspond to models where thermal production leads to the correct DM density today, while upper (lower) triangles indicate a too large (small)  relic density.
The shaded area contains only cMSSM models in the co-annihilation region (see text for further details). Also shown for comparison are the most relevant limits from Tab.~\ref{tab:limits_Majorana}. Right: Same, for neutralino annihilation into $\gamma \gamma$ and $m_\chi\sim130\,{\rm GeV}$.}
\end{figure*}

For illustration, let us have a closer look at a particularly well-motivated example for a Majorana DM candidate, 
the lightest supersymmetric neutralino \cite{Jungman:1995df},
which is  a linear combination of the superpartners of the gauge and Higgs fields,
\be
  \chi\equiv\tilde\chi^0_1= N_{11}\tilde B+N_{12}\tilde W^3 +N_{13}\tilde H_1^0+N_{14}\tilde H_2^0\,.
\ee
For this purpose, we plot in Fig.~\ref{fig:SUSYscan} the ratio $r$ as a function of the loop annihilation rate $\sigma v$ for a large number of supersymmetric models that resulted
from a  scan (for details, see Ref.~\cite{Bringmann:2007nk}) over the parameter space of the cMSSM and a phenomenological MSSM-7. Here, we keep of course only models with the correct neutralino masses to account for the observed 130\,GeV feature. In the figure, we also indicate whether the neutralino is mostly Bino ($Z_{\tilde B}\equiv\left|N_{11}\right|^2>0.9$), Higgsino ($Z_{\tilde H}\equiv\left|N_{13}\right|^2+\left|N_{14}\right|^2>0.9$) or mixed, and whether thermal production leads to the correct relic density.

As a first remark, one can clearly see that it is essentially impossible to explain the required large annihilation rate with 
thermally produced neutralinos. For the case of annihilation into $\gamma Z$ \cite{Ullio:1997ke}, furthermore, we essentially recover our general expectation outlined in the preceding paragraphs: in order for the loop-signal to be large, there must be a sizable contribution $r$ from the imaginary part of the amplitude. This confirms that our limits are as stringent as advertised before.
The largest rates, in particular, are obtained if the neutralino has a considerable Higgsino fraction; in this case the annihilation rate into $\gamma Z$ is dominated by $W$-boson loops  and we include, for comparison, the corresponding limit from Tab.~\ref{tab:limits_Majorana} in  Fig.~\ref{fig:SUSYscan}

\begin{figure}[th!]
 \begin{center}
    \includegraphics[width=\columnwidth]{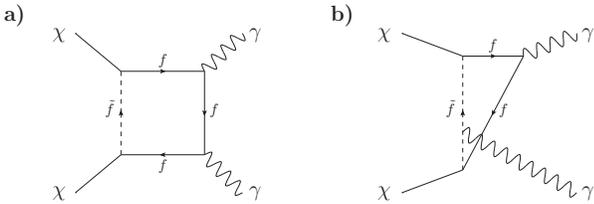} 
 \end{center}
\caption{\label{fig:slepton_box}
For  small mass differences between sfermions $\tilde f$ and Bino-like neutralinos $\chi$, diagram b) completely dominates the process $\chi\chi\rightarrow\gamma\gamma$. This diagram, however,  only has a real part and the largest contribution to the imaginary part of the amplitude derives from diagram a) -- though $W$ boson loops start to contribute with about the same size already for 
 a relatively small Higgsino fraction of the neutralino.
}
\end{figure}

For the $\gamma\gamma$ amplitudes  \cite{Bergstrom:1997fh}, however, there is a large number of models with $\left\langle\sigma v\right\rangle_{\gamma\gamma}\sim 10^{-29}{\rm cm}^3{\rm s}^{-1}$ that do not follow this general expectation and instead show an unexpectedly small value of $r$. While such small annihilation rates are maybe not too relevant in terms of a possible explanation of  the Fermi observation, it is nevertheless quite instructive to discuss the origin of this feature in some more detail. The models in question lie exclusively  in or near the so-called coannihilation region of the cMSSM, where the lightest neutralino is  an essentially pure Bino that is almost degenerate in mass with light sleptons (in particular the $\tilde\tau$). For these models, slepton box diagrams (shown in Fig.~\ref{fig:slepton_box}) dominate the amplitude, with the by far larger contribution coming from  diagram b) that receives a significant enhancement for small mass differences $m_{\tilde \ell}-m_\chi$. This diagram, however, does not have an imaginary part because the virtual leptons cannot simultaneously be put on shell. While this already explains why $r$ should be considerably smaller than expected from our general discussion, there is another effect which can further decrease its value  by up to 4 orders of magnitude or so: in some cases, the imaginary part of the amplitude receives an almost equal contribution from slepton box diagrams (i.e.~the left diagram in Fig.~\ref{fig:slepton_box}) and $W$ boson box diagrams -- albeit with a different sign and thus potentially leading to large accidental cancellations. As indicated in the figure, some of the models in the co-annihilation region therefore evade our bounds on lepton loops even when electrons contribute significantly. The {\it largest} and thus most relevant  annihilation rates, on the other hand, are again obtained for Higgsinos; in this case, the rate is dominated by $W$ boson loops and thus excluded by our constraints (see also Refs.~\cite{Buchmuller:2012rc,Cohen:2012me}).

The above example of neutralino DM nicely illustrates the general caveats one should keep in mind for our analysis: Much smaller values for $r$ than the limits presented in Table \ref{tab:limits_Majorana} can be realized in concrete models if 
\begin{itemize}
\item there are {\it cancellations} between several channels contributing to the imaginary part of the amplitude or
\item the imaginary part for the dominant channel is {\it suppressed} with respect to the real part by some {\it symmetry} of the initial state that can only be satisfied for diagrams without imaginary contributions. In the neutralino case discussed above, e.g., the production of intermediate states in the left diagram of Fig.~\ref{fig:slepton_box} is helicity-suppressed, while this is not the case for the right diagram. 
\end{itemize}
Note that the first point only formally avoids our constraints as there is of course no such cancellation for the tree-level final states and thus for the production of secondary particles. The second point, on the other hand, {\it does} present a fundamental limitation to our analysis as it corresponds to the situation when
 the dominant contribution to the loop signal does not show the same topology as displayed in Fig.~\ref{fig:OT}.

In this particular case,  a sufficiently enhanced  line signal mediated by SM loop particles could thus in principle  explain the observed 130\,GeV feature without violating constraints from tree-level annihilation products. However, it is exactly in this situation that {\it internal bremsstrahlung} (IB) from charged virtual particles generally proceeds at even larger rates.
Majorana DM particles $\chi$, e.g., annihilate with much larger rates into  $\bar f f\gamma$ final states
than into the (helicity-suppressed) $\bar f f$  at tree level \cite{Bergstrom:1989jr} or the ($\alpha_{\rm em}$-suppressed) $\gamma \gamma$/$\gamma Z$ \cite{Bringmann:2007nk}, at least for small mass differences between $\chi$ and $\tilde f$. The crucial point is that this type of 'virtual' IB \cite{Bringmann:2007nk} gives rise to an equally pronounced spectral feature in gamma rays, indistinguishable from a line with the energy resolution of Fermi. It is thus not surprising to see that the ratio of signal to secondary photons can be much larger for IB than for line photons \cite{line_review}, which evades the constraints discussed in this article and may be argued to strengthen the case for a possible IB explanation \cite{Bringmann:2012vr,line_review,Bergstrom:2012bd,Shakya:2012fj} of the signal.

\section{Conclusions}
\label{sec:conc}

The recently found evidence for a line-like spectral feature at 130\,GeV in gamma-ray data towards  the galactic center \cite{Bringmann:2012vr,Weniger:2012tx} has triggered an enormous  activity both in terms of independent data analyses and possible explanations for the observed excess (see Ref.~\cite{line_review} for a review). The most exciting and far-reaching explanation would be in terms of annihilating DM particles $\chi$. In this article, we have studied the possibility that the signal is produced by monochromatic photons resulting from the loop-suppressed process $\chi\chi\rightarrow\gamma\gamma$, $\gamma Z$ or $\gamma h$, and considered possible consequences for DM model-building.

By using the optical theorem, in particular, we could show that the possibility of standard model particles running in those loops is highly constrained by both continuum gamma ray, antiproton and synchrotron cosmic ray data. In fact, it was noted before that the annihilation rate required to explain the 130\,GeV feature is considerably larger than typically expected for WIMPs, so that the associated production of  standard model particles at tree-level would likely  be in conflict with both continuum gamma-ray and antiproton cosmic-ray observations. Here, we have made this argument much more rigorous by adopting a systematic treatment in an as model-independent way as possible.

We note that our results provide important constraints for any model-building attempt to explain the 130\,GeV excess. In particular, a direct coupling between DM and standard model particles (and new, heavier particles) is something very generic in all realistic WIMP frameworks that arise in scenarios that simultaneously try to address the fine-tuning problems of the standard model. In that sense, our findings have far-reaching consequences and are by no means restricted to the case of neutralino DM, which we have explicitly shown to be essentially impossible to reconcile with a monochromatic line interpretation of the data.
While it is in principle possible that the loop signal is dominated exclusively by new particles that are  heavier than the DM particle, such an option may be more difficult to motivate
from a global model-building point of view.

DM annihilation via {\it internal bremsstrahlung}, on the other hand, is a promising alternative explanation: given that it may proceed at much larger rates than the tree-level annihilation into standard model particles, it is  presently not constrained by the data. One should also note that for neutralino DM, e.g., there is only a factor of a few missing in the annihilation rate to explain the observed excess \cite{line_review} -- something which might be possible to achieve by a larger normalization due to a higher local DM density than assumed here \cite{Garbari:2012ff} or an increased DM density near the galactic center due to the presence of baryons, as found in a very recent simulation of a Milky Way-like galaxy \cite{Kuhlen:2012qw}.
 It will thus be exciting to await experiments with improved 
energy resolution and/or statistics,like GAMMA-400 \cite{Galper:2012ji} or HESS-II \cite{hess}, that will be able to distinguish between the peculiar spectrum of internal bremsstrahlung and a monochromatic line \cite{Bergstrom:2012vd}.

\bigskip
{\it Acknowledgements.} 
T.B. and M.A. acknowledge support from the German Research Foundation
(DFG) through the Emmy Noether  grant BR 3954/1-1. M.V.  acknowledges
support from the Forschungs- und Wissenschaftsstiftung Hamburg through
the program ``Astroparticle Physics with Multiple Messengers''.
\bigskip

\appendix


\section{Optical theorem revisited}
\label{app:OT}
In this appendix, we provide for convenience some practical tools for the 
actual  calculation of  the ratio between loop- and 
tree-level cross section. In particular,
we describe two rather different strategies which, however, of course
lead to  consistent results.

\subsection{Standard model rates in Helicity and $(L,S)$ basis}
\label{app:OT1}

The first approach is to actually compute the SM cross section
$\sigma_{f\to I}$ as it appears in Eq.~(\ref{eq:ratios}), where $I$ is the particle species supposed to dominate
the
imaginary part of the loop and
$f=\gamma\gamma,\gamma Z,\gamma h$ is
summed  over all polarization states with the same $J^P$
value as the
initial state of two DM particles (recall that we are only interested in
the limit
of vanishing relative velocity of the DM particles, so that $J$ is simply
given by their  total spin).

This can be done in different ways, for example by expressing $\sigma_{f\to I}$ in terms of final-state plane waves as
\be
 \label{eq:ratios2}
\sigma_{f\to I}\!=\!\frac{\beta_I/\beta_f}{2^{6}\pi^2 s}\sum_{m_{S_f}}\!\!\int\!\! d^2\hat{\bf p}_f|{\cal M}_{(f,\hat{\bf p}_f,S_f,m_{S_f})\to(I,L,S)}(s)|^2\,,
\ee
where the quantum numbers $(J,M,L,S)$ of the intermediate state $I$ are fixed. The amplitudes ${\cal M}_{(f,\hat{\bf p}_f,S_f,m_{S_f})\to(I,L,S)}$ can then be expressed in terms of the plane wave amplitudes ${\cal M}_{(I,\hat{\bf p},S,m_S)\to(f,\hat{\bf p}_f,S_f,m_{S_f})}(s)$ with the aid of Clebsch-Gordan coefficients and spherical harmonics, namely
\begin{widetext}
\begin{eqnarray}\label{eq:ratios3}
{\cal M}_{(f,\hat{\bf p}_f,S_f,m_{S_f})\to(I,L,S)}(s)
&=&  \sum_{m_L,m_S} \int d^2\hat{\bf p}
 Y_{L,m_L}(\hat{\bf p}) \langle J,M | L,m_L,S,m_S\rangle_I
 {\cal M}_{(f,\hat{\bf p}_f,S_f,m_{S_f})\to(I,\hat{\bf p},S,m_S)}(s)
\nonumber \\
&=&  \sum_{m_L,m_{S_i}} \int d^2\hat{\bf p}
 Y_{L,m_L}(\hat{\bf p}) \langle J,M | L,m_L,S,m_S\rangle_I
 \langle S,m_S | S_1,m_{S_1},S_2,m_{S_2}\rangle_I
\nonumber \\
&& \qquad \qquad \times
 {\cal M}_{(f,\hat{\bf
p}_f,S_f,m_{S_f})\to(I,\hat{\bf p},S_1,m_{S_1},S_2,m_{S_2})}(s)
\,,
\end{eqnarray}
\end{widetext}
where we defined the spin and its projection of each intermediate particles as ($S_i, m_{S_i}$) with $i=1,2$.

For $J=0$ (realized for both Majorana and Scalar DM), on the other hand, it  proves useful to consider instead the helicity basis $\{\ket{J,M,\lambda_1,\lambda_2}\}$. In the following, we will
provide details for this approach. Transformation rules between the $(L,S)$ and helicity bases, and a set of interesting properties of the latter, can e.g.~be found in Ref.~\cite{JacobWick:1959}. Following those conventions, in particular, one has
\begin{align}
\label{eq:jacobwick}
&\bracket{J,M,L,S}{J,M,\lambda_1,\lambda_2}=\\
&\qquad\sqrt{\frac{2L+1}{2J+1}} \bracket{J,M=\lambda}{L,m_L\!=\!0,S,m_S\!=\!\lambda} \nonumber\\
&\qquad\times  \bracket{S,m_S\!=\!\lambda}{S_1,m_{S_1}\!=\!\lambda_1,S_2,m_{S_2}\!=\!-\lambda_2}\,, \nonumber
\end{align}
where $\lambda\equiv\lambda_1-\lambda_2$ and the $<...|...>$ on the right-hand side are the standard Clebsch-Gordon coefficients.

\subsubsection{Determination of final and intermediate states}

$J=0$ two-photon states are spanned by just two helicity eigenstates
 $\ket{\!+\!1,+1}$ and $\ket{\!-\!1,\!-\!1}$, with
\bea
\ket{f_S}&\equiv&\frac1{\sqrt 2}\ket{\!+\!1,\!+\!1}+\frac1{\sqrt 2}\ket{\!-\!1,\!-\!1}\\
 \ket{f_M}&\equiv&-\frac1{\sqrt 2}\ket{\!+\!1,\!+\!1}+\frac1{\sqrt 2}\ket{\!-\!1,\!-\!1}
\eea
being the $CP$-even and $CP$-odd eigenstates, respectively. The only allowed two-photon final
state $\ket{f}$ for scalar (Majorana) DM is thus $\ket{f_S}$ ($\ket{f_M}$). The same is true
for $\gamma Z$ in the final state, while $\gamma h$ final states are not possible for $J=0$ (note that the combination $L=1$, $S=1$ cannot be constructed with a massless photon and a scalar particle).

Let us now turn to the possible two-body intermediate states $I$. In case of fermions, there are  only two 
independent helicity eigenstates with $J=0$,  namely $\ket{\!+\!\frac12,+\frac12}$ and $\ket{\!-\!\frac12,\!-\!\frac12}$, and the $CP$ eigenstates coincide with the partial wave basis elements:
\bea
\ket{I_S^{\bar ff}}&\equiv&\ket{L\!=\!1,S\!=\!1}=-\frac1{\sqrt 2}\ket{\!+\!\frac12,+\frac12}-\frac1{\sqrt 2}\ket{\!-\!\frac12,\!-\!\frac12}\,,\nonumber \\ \\
 \ket{I_M^{\bar ff}}&\equiv&\ket{L\!=\!0,S\!=\!0}=\frac1{\sqrt 2}\ket{\!+\!\frac12,+\frac12}-\frac1{\sqrt 2}\ket{\!-\!\frac12,\!-\!\frac12}\,.\nonumber \\
\eea
In the case of two $W$ bosons, on the other hand, there are {\it three} independent states
with $J=0$; one $CP$-odd state (given by  $\ket{L=1,S=1}$) and two $CP$-even states which
can e.g.~be chosen as $\{\ket{L=0,S=0},\ket{L=2,S=2}\}$ or, equivalently, transverse and longitudinal states:
\bea
 \ket{I_S^{WW,t}}&\equiv&\frac1{\sqrt 2}\ket{\!+\!1,\!+\!1}+\frac1{\sqrt 2}\ket{\!-\!1,\!-\!1}\,,\\
\ket{I_S^{WW,l}}&\equiv&\ket{0,0}\,,\\
 \ket{I_M^{WW}}&\equiv&\ket{L\!=\!1,S\!=\!1}=-\frac1{\sqrt 2}\ket{\!+\!1,\!+\!1}+\frac1{\sqrt 2}\ket{\!-\!1,\!-\!1}\,.\nonumber \\
\eea
Which linear combination of $\ket{I_S^{WW,t}}$ and $\ket{I_S^{WW,l}}$ is realized as the dominant contribution to the loop process (in the case of scalar DM) can thus not be determined in terms of the symmetries but depends on the SM and  DM interaction. 

\subsubsection{Computation of helicity amplitudes}

With the above definitions, it is now straight-forward to express any amplitude $\bra IT\ket f$ with $J=0$ 
as a sum of helicity amplitudes $\mathcal{M}^h_{\lambda_\gamma;\lambda_I}\equiv\bra{\lambda_I,\lambda_I}T\ket{\lambda_\gamma,\lambda_\gamma}=\mathcal{M}^h_{-\lambda_\gamma;-\lambda_I}$.
Those $\mathcal{M}^h_{\lambda_\gamma;\lambda_I}$ can also be written in terms of plane waves with spins pointing in the direction of motion, i.e.~the familiar 4-momentum two particle states with definite helicities. For the states, we have
 \cite{JacobWick:1959}
\begin{align}
&\ket{J,M,\lambda_1,\lambda_2}=\\
&\qquad\sqrt{\frac{2J+1}{4\pi}}\int\dif{}^2\hat{\bvec k}\,  d^J_{M,\lambda}(\theta')\e^{-i(M-\lambda)\varphi'} \ket{\hat{\bvec k};\lambda_1,\lambda_2}\,, \nonumber
\end{align}
where again $\lambda\equiv\lambda_1-\lambda_2$,  and $(\theta',\varphi')$ are the angles defining $\hat{\bvec k}$ with respect to some arbitrary cartesian system of coordinates. By noting that $d^0_{0,0}=1$, we may thus write
\bea
\mathcal{M}^h_{\lambda_\gamma;\lambda_I}&=&\int\frac{\dif{}^2\hat{\bvec k}}{\sqrt{4\pi}}\bra{J=0,M=0,\lambda_I,\lambda_I}T\ket{\hat{\bvec k};\lambda_\gamma,\lambda_\gamma} \nonumber\\
 &=& \sqrt{4\pi}\bra{J=0,M=0,\lambda_I,\lambda_I}T\ket{\hat{\bvec z};\lambda_\gamma,\lambda_\gamma} \nonumber\\
 &=&\int\dif{}^2\hat{\bvec p}\bra{\hat{\bvec p};\lambda_I,\lambda_I}T\ket{\hat{\bvec z};\lambda_\gamma,\lambda_\gamma} \nonumber\\
\label{momdec}
 &\equiv& 2\pi\int_0^\pi\mathcal{M}_{\lambda_\gamma;\lambda_I}(\theta)\sin\theta\dif\theta\ ,
\eea
where the second line follows from the spherical symmetry of the state $\ket{J=0,M=0,\lambda_I,\lambda_I}$ and rotational invariance of the interaction, and in the last step we used the fact that the integrand does not depend on the azimuthal angle $\varphi$ ($\bvec z$ is an arbitrary direction, taken to be aligned with the photon momentum,  and $\theta$ is the interaction angle with respect to it, i.e.~the center-of-mass angle between the different momenta/helicities). 

$\mathcal{M}_{\lambda_\gamma;\lambda_I}(\theta)$, as introduced above, is the amplitude for two photons (or one photon and one $Z$ boson) with equal helicities and opposite momenta that annihilate into a pair of charged particles with opposite spins parallel to their momenta. These amplitudes can directly be calculated by standard Feynman rules (see below for our conventions and more details)  and we find
\begin{widetext}
\begin{align}
\label{eq:Amp1}
\mathcal{M}^{\gamma\gamma\to\bar ff}_{\lambda_\gamma;\lambda_{1/2}}(\theta)&=\frac{2(eQ)^2(\beta_I+2\lambda_{1/2}\lambda_\gamma)}{\gamma_I(1-\beta_I^2\cos^2\theta)}\,,\\
\mathcal{M}^{\gamma\gamma\to W^+W^-}_{\lambda_\gamma;\lambda_W}\!\!(\theta)&=\frac{2e^2(2\lambda_W^2+2\lambda_W\lambda_\gamma\beta_I-\gamma_I^{-2})}{(1-\beta_I^2\cos^2\theta)}\, ,\\
\mathcal{M}^{\gamma Z\to\bar ff}_{\lambda_\gamma;\lambda_{1/2}}(\theta)&=\,\frac{geQ}{\gamma_I\beta_f\cos\theta_W}\frac1{1-\beta_I^2\cos^2\theta}\Big{(}C_V^f\left[\beta_I\sin^2\theta+\beta_I\beta_f\cos^2\theta+2\lambda_{1/2}\lambda_\gamma\beta_f\right]+ \nonumber\\
 &  \qquad + C_A^f(\lambda_\gamma\beta_I+2\lambda_{1/2})\beta_f\cos\theta\Big{)}\, ,\\
\mathcal{M}^{\gamma Z\to W^+W^-}_{\lambda_\gamma;\lambda_W\neq0}\!\!(\theta)&=2eg\cos\theta_W\frac{\beta_f(1+\lambda_\gamma\lambda_W\beta_I)^2\cos^2\theta+(\beta_f+\lambda_\gamma\lambda_W\beta_I)^2\sin^2\theta}{\beta_f(1-\beta_I^2\cos^2\theta)}\, ,\\
\label{eq:Amp5}
\mathcal{M}^{\gamma Z\to W^+W^-}_{\lambda_\gamma;\lambda_W=0}\!\!(\theta)&=-2eg\cos\theta_W\frac{\beta_f(1-m_Z^2/2m_W^2)\cos^2\theta +(1-m_Z^2/4m_W^2-m_Z^4\beta_I^2/8m_W^4)\sin^2\theta}{\gamma_I^2\beta_f(1-\beta_I^2\cos^2\theta)}\, ,
\end{align}
\end{widetext}
where the reduced velocity $\beta_I$ of the intermediate state particles $I$ is defined  in Eq.~(\ref{eq:beta_eff}), with a corresponding definition holding for the reduced velocity $\beta_f$ of the final state particles,
and $\gamma_I\equiv\left(1-\beta_I^2\right)^{-1/2}=m_\chi/m_{f(W)}$. 
Furthermore, we introduced $C_V^f\equiv T_3-2Q\sin\theta_W$ and $C_A^f\equiv T_3$, where $Q$ and $T_3$ are the fermion charge and weak isospin projection (for left-handed fermions), respectively.
From these expressions, the helicity amplitudes $\mathcal{M}^h_{\lambda_\gamma;\lambda_I}$ are obtained by integration over $\theta$. 
As explained above, the amplitudes $\bra IT\ket f$ that appear in Eq.~(\ref{eq:ratios}) can then easily be computed as a sum over all contributing $\mathcal{M}^h_{\lambda_\gamma;\lambda_I}$, leading to the results shown in Tab.~\ref{tab:OT_Majorana}.

\subsubsection{Conventions for polarization vectors and spinors}
In arriving at the amplitudes (\ref{eq:Amp1}-\ref{eq:Amp5}), we adopted the following explicit representations 
of the appearing 4-momenta (with $p/E_p\equiv\beta_I$): 
\bea
k_\gamma&=&
  \left(k,0,0, k\right)\,,\\ 
k_{\bar\gamma(Z)}&=&
 \left(E_k, 0, 0, -k \right)\,,\\
p_{I}&=&
 \left(E_p, p\sin\theta, 0, p\cos\theta \right)\,,\\
p_{\bar I}&=&
  \left( E_p, -p\sin\theta, 0, -p\cos\theta\right)\,.
\eea
For fermion states $I$, we used the Weyl representation for  gamma matrices and the following set of spinors:
\begin{align}
&u(\lambda_{1/2}) =\!\! \left(\!
\begin{array}{c}
 \sqrt{E_p-2\lambda_{1/2}p}\cos\!\left(\frac{2\theta+(1-2\lambda_{1/2})\pi}{4}\right) \\
 \sqrt{E_p-2\lambda_{1/2}p}\sin\!\left(\frac{2\theta+(1-2\lambda_{1/2})\pi}{4}\right) \\
 \sqrt{E_p+2\lambda_{1/2}p}\cos\!\left(\frac{2\theta+(1-2\lambda_{1/2})\pi}{4}\right) \\
 \sqrt{E_p+2\lambda_{1/2}p}\sin\!\left(\frac{2\theta+(1-2\lambda_{1/2})\pi}{4}\right)
\end{array}
\!\right)\!,\!\\ 
&v(\text{-}\lambda_{1/2}) =\!\! \left(\!
\begin{array}{c}
\sqrt{E_p-2\lambda_{1/2}p}\sin\!\left(\frac{2\theta-(1-2\lambda_{1/2})\pi}{4}\right) \\
  \sqrt{E_p-2\lambda_{1/2}p}\cos\!\left(\frac{2\theta-(1-2\lambda_{1/2})\pi}{4}\right) \\
  \sqrt{E_p+2\lambda_{1/2}p}\sin\!\left(\frac{2\theta+(3+2\lambda_{1/2})\pi}{4}\right) \\
  \sqrt{E_p+2\lambda_{1/2}p}\cos\!\left(\frac{2\theta+(3+2\lambda_{1/2})\pi}{4}\right) 
\end{array}
\!\right)\!.\!\!\!
\end{align}
Last but not least, the polarization vectors for the vector bosons in our conventions 
read
\begin{align}
\epsilon_\gamma(\lambda_\gamma)&=
\left(0, -{\lambda_\gamma}/{\sqrt 2}, -{i}/{\sqrt 2}, 0\right)\,,\\   
\epsilon_{\bar\gamma(Z)}(\lambda_\gamma)&=
\left(0, {\lambda_\gamma}/{\sqrt 2}, -i/{\sqrt 2}, 0\right) \,,\\
\epsilon^*_W(\lambda_W)&=\left(
\begin{array}{c}
 (1-\lambda_W^2)\gamma_I\beta_I \\
 -\frac{\lambda_W\cos\theta}{\sqrt 2}+(1-\lambda_W^2)\gamma_I\sin\theta \\
 \lambda_W^2\frac{i}{\sqrt 2} \\
 \frac{\lambda_W\sin\theta}{\sqrt 2}+(1-\lambda_W^2)\gamma_I\cos\theta
\end{array}
\right)^T\!,\\
\epsilon^*_{\bar W}(\lambda_W)&=\left(
\begin{array}{c}
 -(1-\lambda_W^2)\gamma_I\beta_I \\
 \frac{\lambda_W\cos\theta}{\sqrt 2}+(1-\lambda_W^2)\gamma_I\sin\theta \\
 \lambda_W^2\frac{i}{\sqrt 2} \\
 -\frac{\lambda_W\sin\theta}{\sqrt 2}+(1-\lambda_W^2)\gamma_I\cos\theta
\end{array}
\right)^T \!.
\end{align}

\subsection{Effective coupling approach}
\label{app:OT2}

The second option (used e.g.~in Ref.~\cite{Abazajian:2011tk}) is to introduce a fictitious particle $\phi$, with $m_\phi=2m_\chi$ 
and the same parity and spin as the initial state. One then needs to specify an
effective 3-point coupling between $\phi$ and the two intermediate states $I$, which 
corresponds to contracting the 'blob' in Fig.~\ref{fig:OT} to a point and considering
the decay of $\phi$ instead of the annihilation of two particles $\chi$. 
While this approach does not explicitly make reference to the general 
formulae presented in Section~\ref{sec:OT}, it is sometimes more convenient to 
follow in practice. Note, in particular, that our general 
discussion in Section~\ref{sec:OT} implies that the {\it ratio} $r_{i\to f}$ between loop- and tree-level cross section, or decay
rate,  does not depend on the form of the effective operator that is chosen (as long as it correctly
connects particles of the required spin and parity).

The reason why this approach can be computationally  simpler is that one can directly
compute the processes $i\rightarrow I$ and $i\rightarrow f$ by standard means, i.e.~by using the full  polarization and/or spin 
sums and without having to worry about explicit projections on specific $(L,S)$ states.
Even the loop integral for $i\rightarrow f$ contains only three legs and is thus relatively 
straight-forward to compute by standard techniques. In fact, one may even directly compute ${\rm Disc}\,\mathcal{M}_{i\rightarrow f}\equiv 2i\Im \mathcal{M}_{i\rightarrow f}$ by using the cutting rules  proven by Cutkosky \cite{Cutkosky:1960sp}: simply take the full expression for $\mathcal{M}_{i\rightarrow f}$ and replace 
\begin{equation}
 \frac{1}{q^2-m_I^2+i\epsilon} \rightarrow -2\pi i\, \delta\!\left(q^2-m_I^2\right)
\end{equation}
for the two propagators of the (necessarily on-shell) intermediate particles. This 'trick'  reduces the loop integral to two angular integrals (one of which is trivial) and makes the integrand much simpler because only one remaining propagator is involved.
Note that the effective operator approach also allows to rather easily calculate the contribution to the imaginary part of the loop integral that derives from cutting along the dotted line of Fig.~\ref{fig:OT}, which appears only for $\gamma Z$ final states dominated by fermion loops and which is neglected in Eq.~(\ref{eq:ratios}) and thus in the approach described in the previous subsection.

For completeness, let us finally list all  effective operators that are needed to compute the relevant rates for annihilating 
Majorana and scalar DM. In the Majorana DM case, $\phi$ is a pseudoscalar that interacts with fermions via 
\be
 \mathcal{L}^{\rm int}_{\phi\bar f f}=i g_{\phi_M,f}\,\phi\bar f\gamma^5 f
\ee
and with gauge bosons via
\bea
 \mathcal{L}^{\rm int}_{\phi WW}&=&g_{\phi_M,W}\,\phi\, (\partial_\mu W^+_\nu) (\partial_\rho W^-_\sigma) \epsilon^{\mu\nu\rho\sigma}\\
 &\subset&g_{\phi_M,2}\,\phi F^a_{\mu\nu}\tilde F_a^{\mu\nu}\,,
\eea
where
 $F^a_{\mu\nu}$ denotes the $SU(2)$ field strength.
In the scalar DM case, on the other hand, $\phi$ is a scalar that interacts with fermions via
\be
 \mathcal{L}^{\rm int}_{\phi\bar f f}=g_{\phi_S,f}\,\phi\bar f f
\ee
and with gauge bosons via
\bea
 \mathcal{L}^{\rm int}_{\phi WW}&=&g_{\phi_S,W}\,\phi\, W^+_\mu\left(\partial^\mu \partial^\nu - \eta^{\mu\nu} \partial^2 \right) W^-_\nu \\ 
 &\subset&g_{\phi_S,2}\,\phi F^a_{\mu\nu} F_a^{\mu\nu}\,.
\eea
As stressed above, the coupling constants $g_{\phi,...}$ that appear in these expressions drop out when calculating ratios like $r$; alternatively, they may be fixed by normalizing to either the tree or loop level rate.

\end{document}